\documentclass[aps,reprint,superscriptaddress,amsmath,amssymb,showpacs,floatfix]{revtex4-2}
\usepackage{graphicx}
\usepackage{bm}
\usepackage{color}
\usepackage{lipsum}

\usepackage{textcomp}
\usepackage{mathptmx}
\usepackage[colorlinks=true,allcolors=blue,urlcolor=black]{hyperref}

\usepackage{xr}

\makeatletter
\newcommand*{\addFileDependency}[1]{
	\typeout{(#1)}
	\@addtofilelist{#1}
	\IfFileExists{#1}{}{\typeout{No file #1.}}
}
\makeatother

\begin{document}
\title{Accessing power-law statistics under experimental constraints}
\author{Xavier Durang}
\email[]{xdurang@postech.ac.kr}
\affiliation{Department of Physics, Pohang University of Science and Technology, Pohang 37673, Republic of Korea}
\author{Hyerim Ahn}
\affiliation{Department of Physics and Astronomy, Seoul National University, Seoul 08826, Republic of Korea}
\affiliation{Department of Electrical and Computer Engineering, University of Minnesota, Minneapolis 55455, USA}
\author{Jae Youn Shim}
\affiliation{Department of Physics and Astronomy, Seoul National University, Seoul 08826, Republic of Korea}
\author{Hye Yoon Park}
\email[]{hyp@umn.edu}
\affiliation{Department of Physics and Astronomy, Seoul National University, Seoul 08826, Republic of Korea}
\affiliation{Department of Electrical and Computer Engineering, University of Minnesota, Minneapolis 55455, USA}
\author{Jae-Hyung Jeon}
\email[]{jeonjh@postech.ac.kr}
\affiliation{Department of Physics, Pohang University of Science and Technology, Pohang 37673, Republic of Korea}
\affiliation{Asia Pacific Center for Theoretical Physics, Pohang 37673, Republic of Korea}

\date{\today}

\begin{abstract}


Over the last decades, impressive progresses have been made in many experimental domains, e.g. microscopic techniques such as single-particle tracking, leading to plethoric amounts of data. In a large variety of systems, from natural to socio-economic, the analysis of these experimental data conducted us to conclude about the omnipresence of power-laws. For example, in living systems, we are used to observing anomalous diffusion, e.g. in the motion of proteins within the cell. However, estimating the power-law exponents is challenging. Both technical constraints and experimental limitations affect the statistics of observed data. Here, we investigate in detail the influence of two essential constraints in the experiment, namely, the temporal-spatial resolution and the time-window of the experiment. We study how the observed distribution of an observable is modified by them and analytically derive the expression of the power-law distribution for the observed distribution through the scope of the experiment. We also apply our results on data from an experimental study of the transport of mRNA-protein complexes along dendrites.

\end{abstract}

\maketitle
\section{Introduction}
Found in critical phenomena \cite{Kada1967,Ashl1971,Lipa1996,Sorn2004}, in systems with a critical self-organization \cite{Bak1987}, earthquake magnitudes \cite{Nekr2020}, human mobility \cite{Zhao2015}, animal foraging or distribution pattern of animal species \cite{Jame2011}, and transport in cells \cite{Chen2015, Song2018}, the ubiquity of power-laws in natural, technical and living systems is evident and has attracted tremendous research activities to understand their origins. 
Typical experimental data consist of either time series of events or trajectories of specific objects. In the former case, the events are recorded along the timeline (Fig.~ \ref{fig1}(a)), e.g. earthquake sequences \cite{Jons2006}, and the inter-event time distribution is usually studied. In the later case, a prototypical example is the single-particle tracking experiment which is routinely used to record the movement of labeled macromolecules within their native environment, e.g. \cite{Song2007, Harr2012, Song2018, Koni2020, Bayl2021}. 

Indeed, over the last decades, a huge array of experimental techniques and analytical tools, such as single-particle tracking technique, has been applied to characterize the microscopic behavior in the living systems. From the measured trajectories, the diffusion coefficient or the actual transport properties are deduced and microscopic models mimicking the observed behavior are constructed. For example, the motion of $\beta$-actin mRNA-protein complex was shown to follow an aging L\'evy walk \cite{Song2018}. However, the estimation of the power-law exponent appearing in those experiments is nontrivial, where experimental constraints play a critical role in its determination.

The anomalous diffusion of a single particle is conventionally classified by a power-law scaling of the mean-squared displacement:
\begin{equation}\label{eq1}
\langle x^2(t) \rangle \propto t^\alpha  
\end{equation}
where $\alpha\neq1$ is referred to as anomalous exponent. The diffusion process is called subdiffusion for $0<\alpha<1$ and superdiffusion for $\alpha>1$~\cite{metzler2014review}. 
This power-law exponent appears in various areas such as target finding times \cite{Metz2009} or cellular organization \cite{Bark2012}. The generalized diffusion law \eqref{eq1} emerges due to the breakdown of the central limit theorem (CLT). One of the physical mechanisms violating the CLT is broad distributions in diffusion events, such as jump lengths or waiting times between successive jumps~\cite{metzler2014review}. It is known that anomalous diffusion processes in this category are described by the diffusion models in the class of continuous-time random walk (CTRW) such as subdiffusive CTRW (Fig.~\ref{fig1}(b)), L\'evy walk (LW), and LW with rests (Fig.~\ref{fig1}(c)). They can be understood as a two-state process where the system stays in a state for a duration drawn from a given distribution; see the rest events in the subdiffusive CTRW (Fig.~\ref{fig1}(b)) and run/rest events in the LW with rests (Fig.~\ref{fig1}(c)). For many single-particle tracking experiments, it has been reported that these models describe the microscopic transport dynamics successfully. Examples include the gamma burst pattern in a primate cerebral cortex \cite{Liu2021}, the motion of mRNA along the dendrites \cite{Song2018}, the diffusion of microbeads in cytoskelectal filaments~\cite{wong2004PRL}, the predator search behavior \cite{Sims2008}, the human mobility \cite{Rhee2011}, the migration of swarming bacteria \cite{Arie2015}, the central pattern of locomotion of the Drosophila~\cite{Sims2019}, and the T-cell motility in the brain \cite{Harr2012}. 

In the aforementioned dynamic models such as time series of events or CTRW families, the knowledge on the distribution or its probability density function (PDF) of random event times is essential to classify the dynamics of the object.
However, it is highly nontrivial to correctly obtain the corresponding time (or length) distribution in the experiments due to the limitations and errors of the measurement. For instance, in single-particle tracking experiments, the localization errors that originate from photon-counting noise, pixelation noise, and background noise \cite{Thom2002,bobr1986, Savi2005, Berg2010}, have been shown to induce a bias in the determination of the power-law (or anomalous) exponent \cite{Mart2002}. This effect cannot be fully resolved by an improved ensemble average or a longer experiment  \cite{Kept2013,Burn2015}. The experimental errors can be viewed to introduce some noise that hinders the determination of the state in which the system is. A minimal number of successive measurements is therefore required to assign a state to the system. Furthermore, the duration of the experiments itself affects the apparent distribution of step size: it is usually treated using an upper truncated Pareto distribution and has been applied in different contexts, e.g. in finance \cite{Aban2006,Sagi2013,Shar2021}, geology \cite{Scho2012}, and biology \cite{Sims2019}. In this work, we show how to take into account these two constraints in order to calculate and recover the intrinsic distributions of the duration of the states of the system.

The organization of the paper is in the following. In Sec.~II, we define the dynamic processes under consideration with three distinct dynamic models (Fig.~\ref{fig1}) and establish a mathematical framework. In Sec.~\ref{secIIa}, we construct a theory dealing with the temporal-spatial resolution effect. In Sec.~\ref{secIIb}, we investigate the modification of event time PDFs due to the effect of a finite observation window. Thereafter, in Sec.~\ref{secIIc} we combine the two effects, constructing the complete theory for accessing the event time PDF under the limits of the resolution and finite observation time window.  
For each case, we provide a formal expression of the observed PDF of the duration/length of events and give an analytical expression for the special case of the power-law distribution. For confirming the pertinence of our theory, we also perform simulations of the CTRW and LW models and fit the distribution of the duration of events with these analytical expressions. In Sec.~\ref{secIII}, we apply our theory to an experimental case: the transport of mRNA-protein (mRNP) particles along dendrites in neuronal cells. We extract the power-law exponent of the distribution of the duration of the immobile state of these macromolecular complexes and discuss the results therein. Finally, we give some concluding remarks.

\section{Theory}
\begin{figure}
\includegraphics[width=0.9\linewidth]{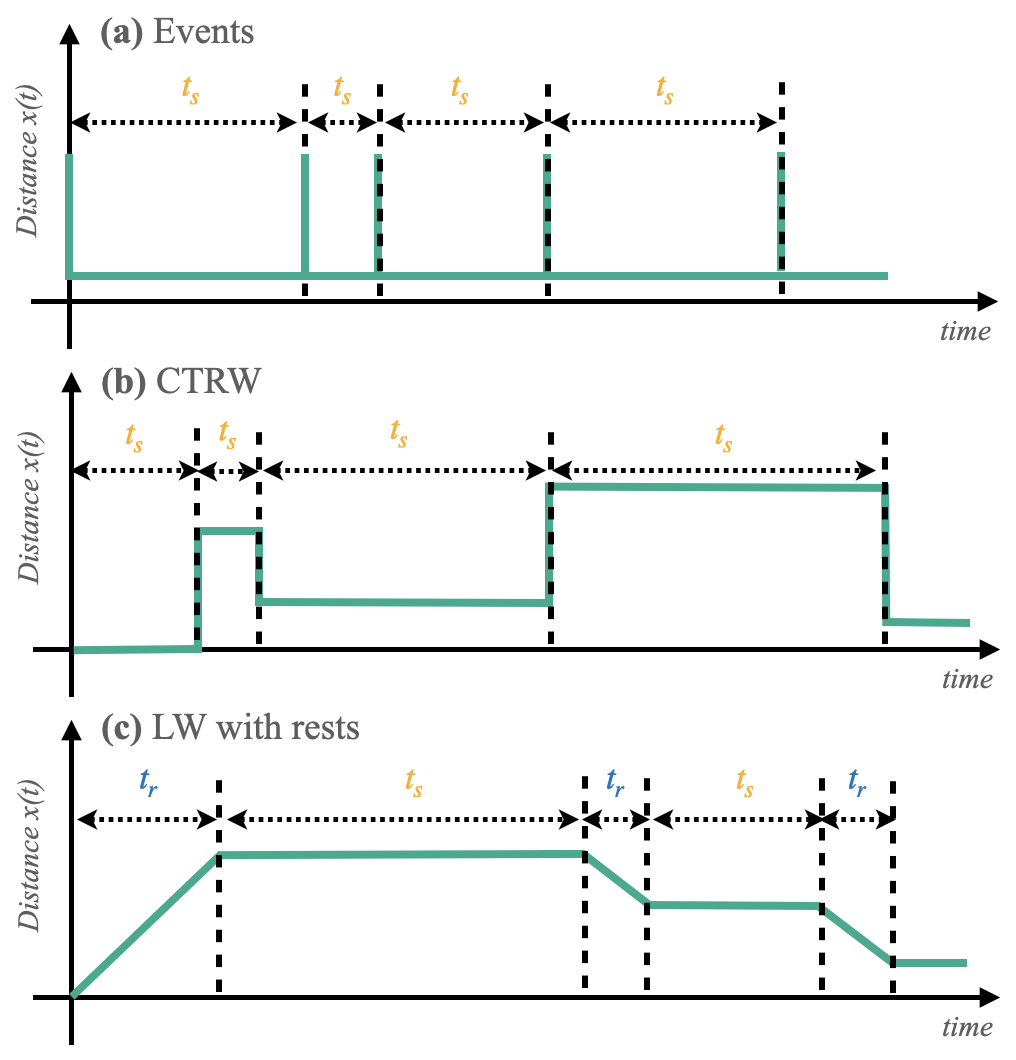}
\caption{Examples of stochastic processes under consideration. (a) Events occurring along the timeline, e.g. earthquakes. The quantity of interest is the PDF of the inter-event time $t_s$. (b) Continuous-time random walk processes governed by $\psi_{\rm waiting}(t_s)$, i.e., the PDF of waiting times $t_s$ between successive jumps. The jumps are instantaneous and described by the PDF of jump length $\psi_\mathrm{jump}(x)$. (c) A L\'evy walk with rests~\cite{zaburdaev2015levy}. It is a two-state process consisting of an alternation of a ballistically moving run phase and a "no-moving" rest phase. The occurrence times of both phases are independent, characterized by the corresponding PDFs, 
$\psi_{\rm run}(t_r)$ and $\psi_{\rm rest}(t_s)$, respectively. }
\label{fig1}
\end{figure}

The underlying process we consider is a two-state process  (state $S$ and state $S'$): a particle alternates between one state to the other with a duration drawn from a given PDF $\psi_S(t)$ (resp. $\psi_{S'}(t)$) or with a given rate. 
Representative examples include time series of events, such as earthquake occurrence or arrival of emails (Fig.~\ref{fig1}(a)), and on-off processes like blinking of quantum dots \cite{Efro2016}. In the diffusion process, the continuous-time random walk is a typical example. In the framework of CTRW, a random walk is described by instantaneous jumps and waiting events in between the successive jumps (Fig.~\ref{fig1}(b)). An additional important example is a LW with rests (Fig.~\ref{fig1}(c)). In this process, the instantaneous jump in the (subdiffusive) CTRW is replaced by a ballistic run whose duration time is proportional to the jump length. The process is thus understood as ballistic diffusion alternating with immobile motions. The statistical property of the PDF of the duration time of each state is a fundamental characteristic of the process. 

In practice, the time series data consist of a succession of positions measured every time interval $\Delta t$. Therefore, the data set is given by ${\cal X}=\{ X_i, i=0\ldots N \} $ where $i$ is the time index and $N$ is the maximum number of measurement. A conventional method to identify the state is to calculate the instantaneous velocity $v_{i,{\rm inst}}=\frac{X_{i+1}-X_i}{\Delta t}$. If the instantaneous velocity is below a given threshold, the particle is in an immobile state and is in a mobile state otherwise. This strategy is a projection from the position to the state of the system, and it generates a time series $\{S_i, i=1\ldots N-1\}$ where $S_i$ is the state at time $i\Delta t$. In general, there exist other complicated strategies identifying the system state \cite{Flom2006,Mcki2006,Whit2020}, but they all consist of a projection $\cal F$ from the data set $\cal X$ and result in a time series of states ${\cal S}={\cal F (X)} = \{S_i, i=1\ldots N-r\}$ where $r$ is the minimum data points needed to determine the state of the system. Once ${\cal S}$ is obtained, the duration time of each state and their PDFs are immediately extracted from it.
Our goal in this work is to extract the parameters of the original PDF (especially a power-law PDF) from the data set. We show how the power-law involved data is severely tempered by two effects: the temporal-spatial resolution and time window.


\subsection{Temporal-spatial resolution}\label{secIIa}
Determining the state $\{S_i\}$ requires a minimum number of data points. If one uses the instantaneous velocity aforementioned, it only requires two data points. However, in general it requires more data points because of the noise or of the measurement errors. Those minimum data points are the resolution $r$. In other words, with the knowledge of $r$ data points, we determine the system state. However, in doing so we have implicitly assumed that the state remains unchanged during $r\Delta t$, which is not generally true. There exists a possibility that a change of state occurs during the time interval of $r\Delta t$. Suppose that we have observed that the state $S$ of the system lasts for a certain period of time $t$. There is a probability $P_r$ that during the observation time $t(>0)$ the state has quickly changed to $S'$ during a time interval shorter than $r\Delta t$ at time $\tau_1(<t)$ and come back to $S$. Generalizing this idea, there is a probability $(P_r)^m$ that during the observation time $t$ the state has briefly changed $m$ times to $S'$ for a time period shorter than $r\Delta t$. Assuming that the duration of the state $S'$ are infinitesimal (as usually considered in the time series of the event occurrence and CTRWs in Fig.~\ref{fig1}), the apparent PDF of duration time of the state $S$ can be written in terms of the true PDF $\psi_S(t)$ and $P_r$ as in the following:
\begin{eqnarray}
\label{eq_reso1}
&&\psi_{{\rm reso},S}(t)= \frac{1}{\cal N}\left[\psi_{ S}(t) + P_r\int_0^t {\rm d}\tau_1 \psi_{S}(\tau_1)\psi_{S}(t-\tau_1) \right.\\ \nonumber
&+&\left. {P_r}^2\int_0^t d\tau_1 \psi_S(\tau_1) \int_{\tau_1}^{t} d\tau_2 \psi_S(\tau_2-\tau_1)\psi_S(t-\tau_2) +...\right].
\end{eqnarray}
In R.H.S, the first term takes into account the probability that the state $S$ lasts continuously up to time $t$, the second term that the state $S$ changes briefly to $S'$ at time $\tau_1$, the third that the state $S$ changes briefly to $S'$ two times at $\tau_1$ and $\tau_2$, etc. 


\begin{figure*}
\centering
\includegraphics[width=0.99\linewidth]{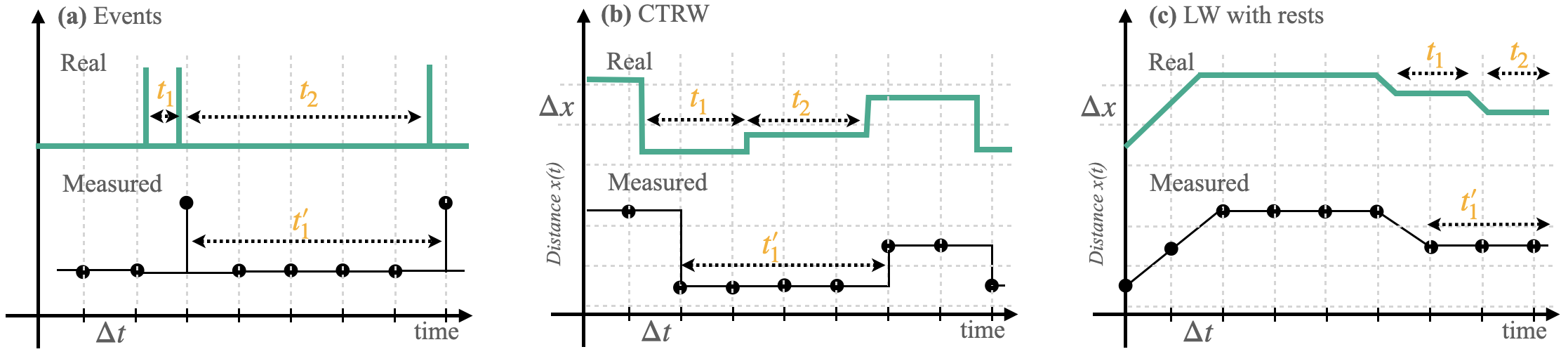}
\par\vspace{0.3cm}
\includegraphics[width=0.24\linewidth]{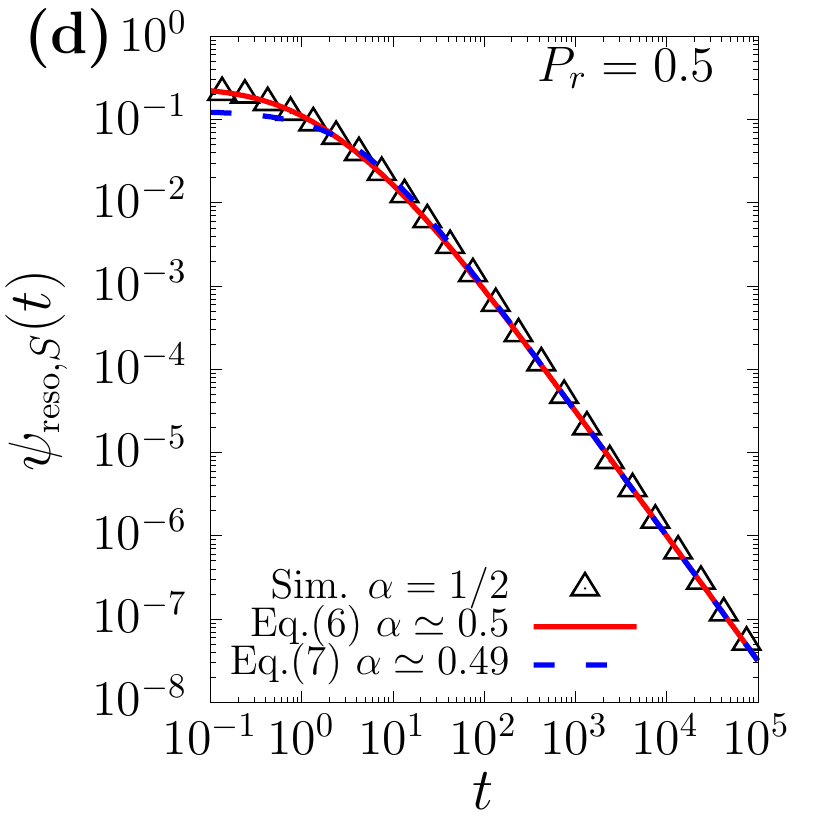}
\includegraphics[width=0.24\linewidth]{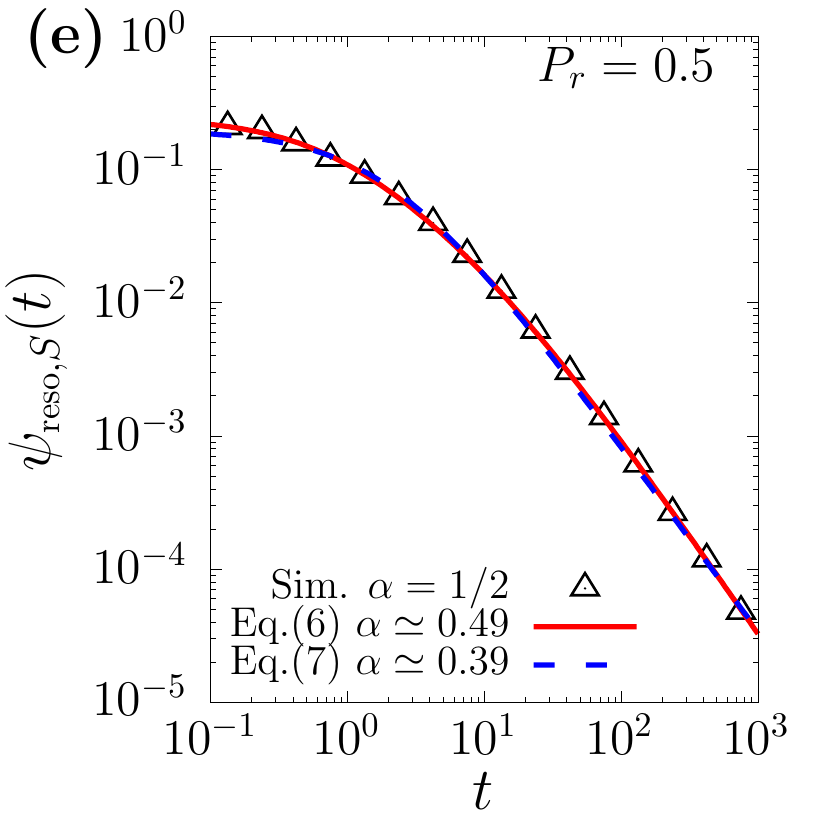}
\includegraphics[width=0.24\linewidth]{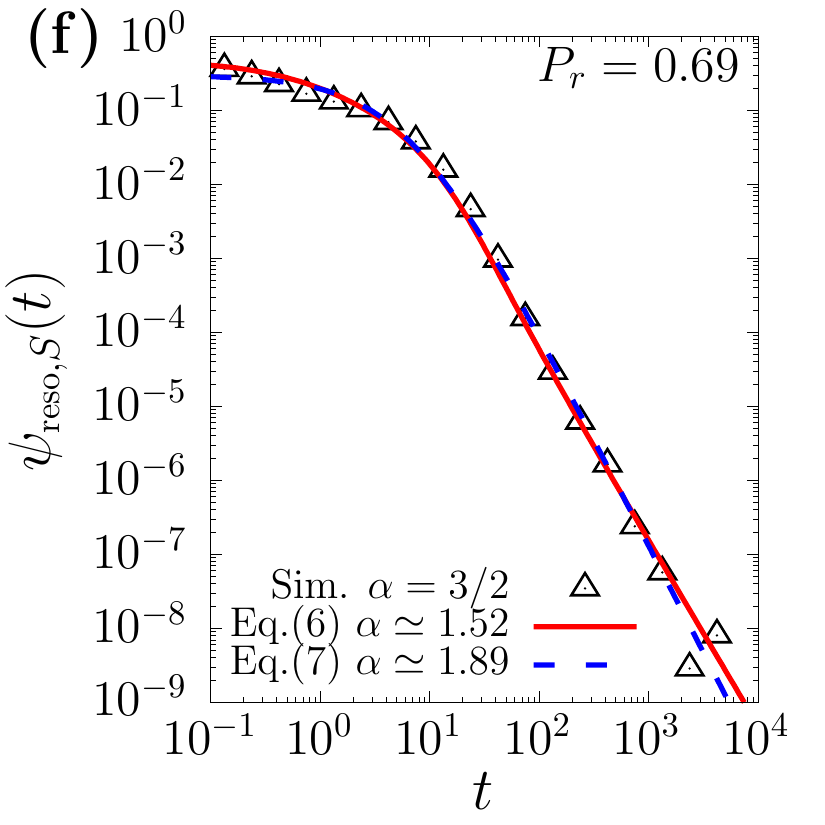}
\includegraphics[width=0.24\linewidth]{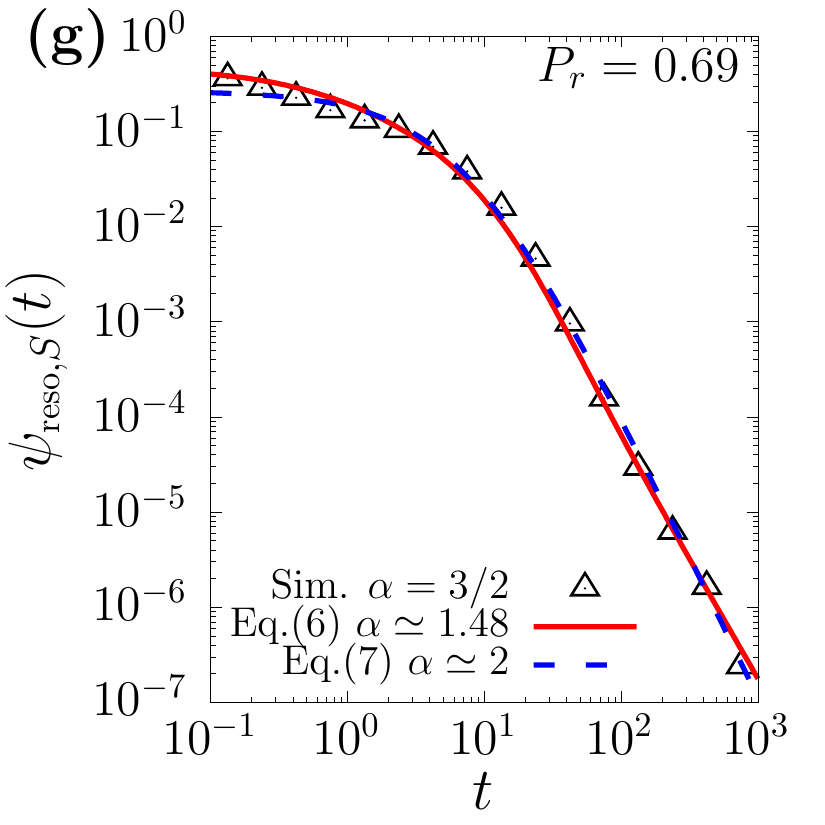}
\caption{Top: The schematics of the measurement process for (a) a time series of events, (b) a CTRW process and (c) a LW with rests; If a change occurs within the time resolution $\Delta t$ and/or the spatial resolution $\Delta x$, the event is undetected. Because of this effect, the statistics of inter-event times or waiting times can be altered substantially. In the panel (a)--(c), we schematically draw the original time sequences of events (upper) and the recorded time sequences under the resolution limit (bottom). 
Bottom: Apparent PDFs $\psi_{\mathrm{reso},S}(t)$ of the waiting times of a CTRW process recorded with a time interval $\Delta t=0.2$ and spatial resolution $\Delta x=0.2$. The CTRW was simulated with the waiting time PDF \eqref{eq:power-law} and the jump length distribution $\psi_{\rm jump}(x)\sim \mathrm{e}^{-x/x_0}$. The followings are the parameters used in the simulation: (d, e) $\alpha=1/2$, $\tau_0=1$,   $x_0=1.44$, and $r\Delta x = 1$ ($P_r=0.5$). (f, g) $\alpha=3/2$, $\tau_0=1$, $x_0=1.71$, and $r\Delta x = 2$ ($P_r=0.69$).
The simulation data were fitted with our theory Eq.~(\ref{eq_reso2}) using the numerical inverse Laplace transform (solid lines) and with the original power-law PDF \eqref{eq:power-law} used in the simulation (dashed lines). The corresponding best fit curve and the fit values are shown in the figure. 
}
\label{fig2}
\end{figure*}

Because $P_r$ is the probability that the state lasts for a period of time shorter than the resolution $r\Delta t$, it depends on which underlying process is considered. For the CTRW, it is the probability that a jump is smaller than the spatial resolution $r\Delta x$
\begin{equation}\label{eq:P_rCTRW}
 P_r=\int_0^{r\Delta x} {\rm d}x\; \psi_{\rm jump}(x)
\end{equation}
where $\psi_{\rm jump}(x)$ is the jump length PDF. For the LW with rests (Fig.~\ref{fig1}(c)), it is the probability that the run is shorter than the temporal resolution
\begin{eqnarray}
\label{eq3}
P_r=\int_0^{r\Delta t} {\rm d}\tau\; \psi_{S'}(\tau)
\end{eqnarray}
where $\psi_{S'}(\tau)$ is the PDF of duration time of run events ($S'$). Thus, $P_r$ is related to the cumulative distribution of the $S'$ state.  



Using a Laplace transform $\hat{f}(s)=\mathcal{L}[f(t)]=\int_0^\infty {\rm d}t f(t){\rm e}^{-st}$, we rewrite Eq.~(\ref{eq_reso1}) as 
\begin{eqnarray}
\hat{\psi}_{{\rm reso},S}(s)&=&\frac{1}{\cal N} \left[\hat{\psi}_{S}(s)+P_r\hat{\psi}_{S}(s)^2+P_r^2\hat{\psi}_{S}(s)^3+...\right]\nonumber\\
&=& \frac{1}{\cal N} \sum_{n=0}^{\infty} P_r^n\hat{\psi}_{S}(s)^{n+1}.
\end{eqnarray}
Because $\psi_S(t)$ is a probability distribution, it ensures $\hat{\psi_S}(s=0)=1$ and the normalization factor $\cal N$ is ${\cal N} = \frac{1}{1-P_r}$. The above geometric series then leads to
\begin{equation}
\label{eq_reso2}
\hat{\psi}_{{\rm reso},S}(s)=\frac{(1-P_r)\hat{\psi}_{S}(s)}{1-P_r\hat{\psi}_{S}(s)}.
\end{equation}
Analytically, we obtain the formal expression for the apparent PDF such that $\psi_\mathrm{reso,S}(t)=\mathcal{L}^{-1}[\hat{\psi}_{{\rm reso},S}(s)]$.
Although performing the inverse Laplace transform is generally not feasible, we are able to evaluate it numerically with the Gaver-Stehfest algorithm \cite{Steh1970} that has been proven to converge exponentially fast if the function is analytic around the point of evaluation \cite{Kuzn2013}. Then, Eq.~(\ref{eq_reso2}) can be used to fit the data and infer the parameters of the true PDF $\psi_\mathrm{S}(t)$.

To investigate the effect of the temporal-spatial resolution, let us consider a normalized power-law PDF
\begin{equation}\label{eq:power-law}
\psi_\mathrm{S}(t)=\frac{\alpha \tau_0^\alpha}{(t+\tau_0)^{1+\alpha}}
\end{equation}
and numerically obtain the event statistics limited by a given resolution. 
Before the numerical study, we discuss the asymptotic behaviors of $\psi_\mathrm{reso,S}(t)$ at the two limiting conditions of $t\to0$ and $t\to\infty$ by the virtue of the Tauberian theorems: (1) The information of ${\psi_{{\rm reso},S}}(t\to0)$ is obtained via  $\lim_{s\rightarrow \infty} \hat{\psi}_{{\rm reso},S}(s) = \frac{\alpha(1-P_r)}{s}$, which gives the relation
\begin{equation}
{\psi_{{\rm reso},S}}(0) = (1-P_r)\psi_S(0).
\end{equation}
A simple interpretation is that the shortest durations are likely to  aggregate into longer events but are less likely to be the results of aggregation themselves.
(2) For the large-time limit and with $\tau_0=1$, we find that $\hat{\psi}_{{\rm reso},S}(s\to0)\simeq 1 - \frac{\alpha s^\alpha}{1-P_r}$, which translates to in the time domain
\begin{equation}
\psi_{{\rm reso},S}(t) \sim \frac{1}{(1-P_r)t^{1+\alpha}}.
\end{equation}
This result suggests that if it is possible to observe a time series within a sufficiently long observation time ($t\to\infty$) one can obtain the power-law exponent of the original PDF by measuring the long time behavior even under the temporal-spatial resolution limit with a deceased amplitude by the factor of $1-P_r$. However, estimating $\alpha$ by means of the true PDF \eqref{eq:power-law} often results in an incorrect value unless a time series is very long, which is tested below. 

In Figs.~\ref{fig2}(d)--(g), we simulated CTRW processes with the condition of $\alpha=1/2$ ($P_r=0.5$) and $\alpha=3/2$ ($P_r=0.69$) for several observation times. Figs.~\ref{fig2}(d) \& (e) shows the apparent PDFs $\psi_{\mathrm{reso},S}(t)$ from the simulation (upper triangle) and two fitted PDFs with our theory Eq.~\eqref{eq_reso2} and the bare PDF Eq.~\eqref{eq:power-law}. Both fittings correctly estimate $\alpha$ with a six-decade long waiting time PDF (Fig.~\ref{fig2}(d)). Note that with a four-decade long statistics the usual fitting method with Eq.~\eqref{eq:power-law} produces a large error in the estimation of $\alpha$ while our theory based on Eq.~\eqref{eq_reso2} works well in this case. 

The resolution effect becomes stronger if the tail of power-law distributions is shorter. See the case of $\alpha=3/2$ presented in Figs.~\ref{fig2}(f)\&(g). The fitting with a power-law PDF \eqref{eq:power-law} seemingly explains well the data. But the fitted values severely deviate from $3/2$ even with a sufficiently large data set.  
On the contrary, we confirm that the estimated values of $\alpha$ with our theory Eq.~\eqref{eq_reso2} are in good agreement with the input value regardless of the observation window $t$ of the data. 

\subsection{Time window}\label{secIIb}
To measure the duration time of a state $S$, an event should occur within the time window $[0,T]$ of the experiment. Thus, the time window will strongly affect the measured PDF as no event longer than the time window itself will be registered; when a duration time PDF has a characteristic timescale such as an exponential distribution, a proper choice of the time window could constitute a solution. However, if tuning a time window is not possible or if the PDF has no characteristic timescale as for power-law distributions, the time window effect has to be taken into account explicitly.
\begin{figure}
\centering
\includegraphics[width=0.9\linewidth]{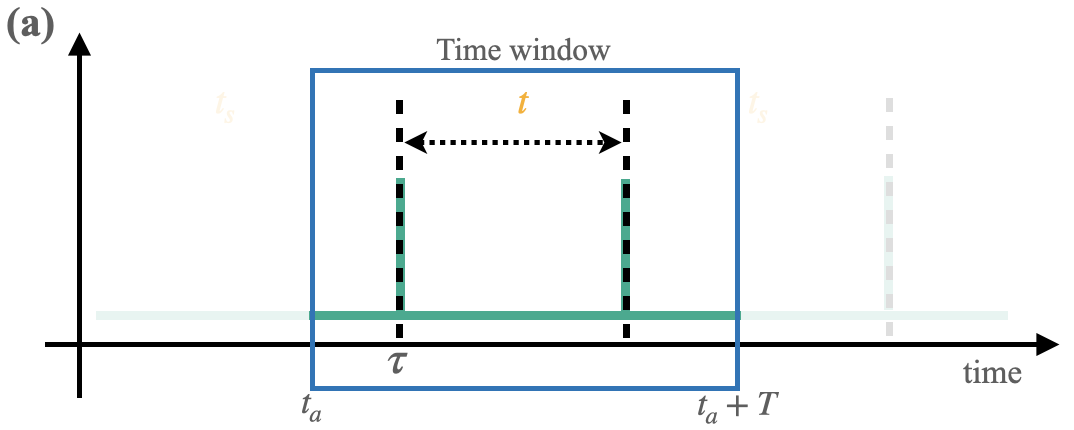}
\par\vspace{0.3cm}
\includegraphics[width=0.9\linewidth]{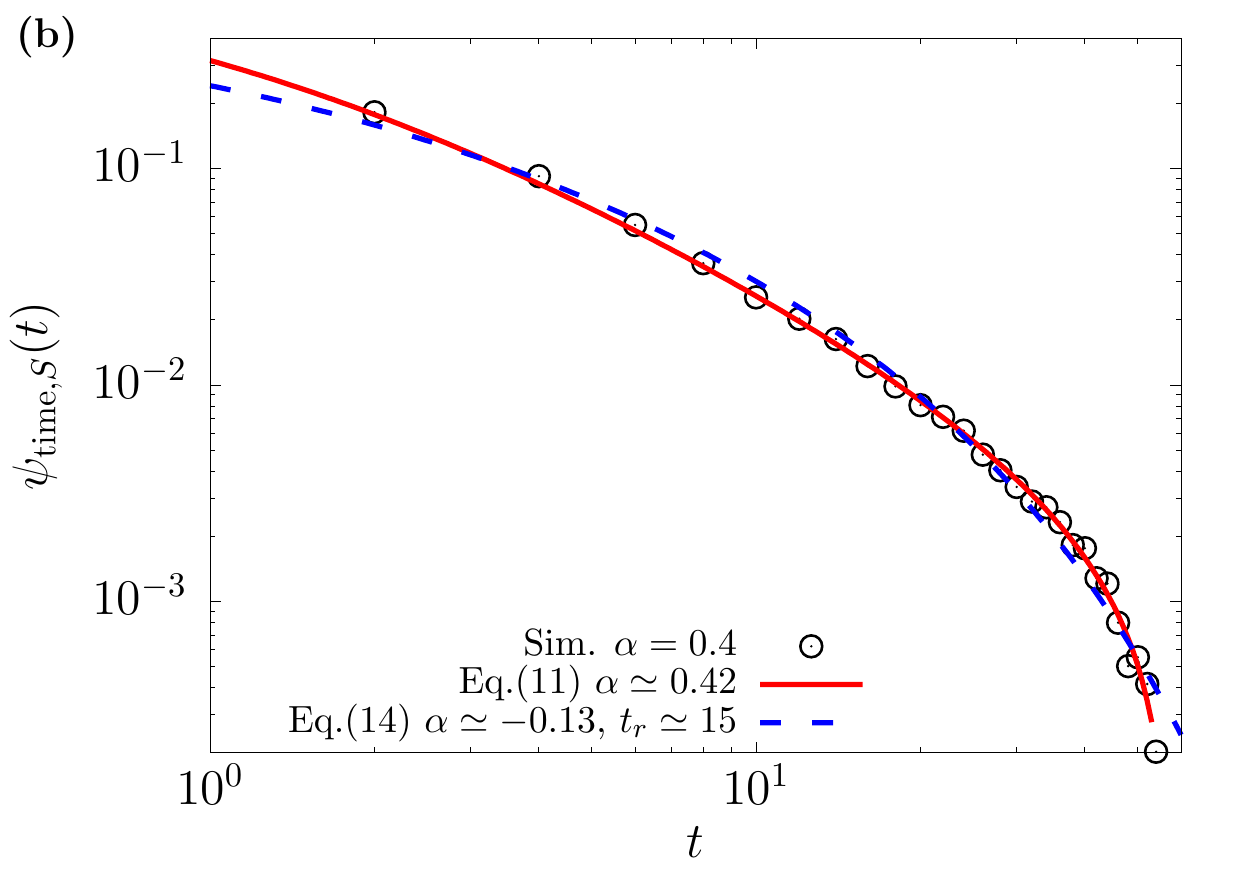}
\caption{(a) The scheme of a typical experiment: $t_a$ represents the beginning of the experiment where the measurements start. $T$ is the duration of the experiment. An event occurs at a random time $\tau$ followed by another event at time $\tau+t$. The inter-event time $t$ is drawn from a given distribution $\psi_S(t)$. 
(b) The waiting time PDF of a CTRW process recorded within a finite observation time window $T$. The data from simulations (symbols) is compared with the fit by our theory (solid line) Eq.~(\ref{eq_timewindow2}) of $\psi_{{\rm time},S}(t)$ with $\tau_0=1$ and by an exponentially truncated power-law Eq.~(\ref{ETPL}) (dashed line). The CTRW was simulated with the waiting time PDF \eqref{eq:power-law} with  $\alpha=0.4$ and $\tau_0=1$. The observation time window was $T=60$. }
\label{fig3}
\end{figure}

In Fig.~\ref{fig3}a, we schematically depict a typical situation to be considered. Assume that during our observation in $[0,T]$ an event occurs at time $\tau$ with a probability $P(\tau)$. The event will last for a duration $t$ distributed according to $\psi_{S}(t)$ (or $\psi_{S'}(t)$). If $t+\tau$ is larger than the time window $T$, then it cannot be registered in the statistics. For a given time $\tau$ for an event to start, the probability of observing the complete event is ${\rm Prob}(t<T-\tau)=\int_0^{T-\tau}{\rm d}u \psi_S(u)$. Therefore, each event registered in the distribution will be weighted by this probability 
\begin{equation}
\label{eq_timewindow}
\psi_{{\rm time},S}(t) = \frac{\psi_S(t)}{\cal N}\int_0^{T-t}d\tau \;P(\tau) \int_0^{T-\tau} du \; \psi_S(u) \Theta(T-t)
\end{equation}
where the step function $\Theta(T-t)$ ensures $t<T$.
Assuming the process is almost time-translation invariant, the probability $P(\tau)$ becomes a constant, absorbed into the normalization. This is typically the case in L\'evy walks when the aging time becomes large enough \cite{Magd2017}. Evaluating the integrals for the power law case Eq.~\eqref{eq:power-law} \cite{Prud1986}, we obtain for $t\in [0,T]$
\begin{eqnarray}
\label{eq_timewindow2}
\psi_{{\rm time},S}(t)&=&\frac{\psi_S(t)}{\cal N}\left[T-t+\tau_0^\alpha\frac{(T+\tau_0)^{1-\alpha}-(t+\tau_0)^{1-\alpha}}{\alpha-1}\right] \nonumber \\
&=& W(t,T)\psi_S(t)
\end{eqnarray}
We note that, contrary to a naive expectation, $\psi_{{\rm time},S}(t)$ is not an exponentially truncated power-law which behaves as $\psi_{{\rm time},S}(t)\simeq \exp(- C\times t/T)\psi_S(t)$. The truncation factor behaves as 
\begin{equation}
\label{eq_TW_shorttime}
W(t,T)\simeq\frac{1}{\cal N}\left[\left(T+\frac{(T+1)^{1-\alpha}\tau_0^\alpha-\tau_0}{\alpha-1}\right)-\frac{\alpha}{2\tau_0}t^2\right]
\end{equation}
when $t$ is small. This means that $\psi_{{\rm time},S}(t)\propto \psi_S(t)$ up to a prefactor. When $t\to T$, the truncation factor is approximated to 
\begin{equation}
W(t,T)\simeq\frac{1}{\cal N}\left[1-\frac{\tau_0^\alpha}{(T+\tau_0)^{\alpha}}\right](T-t).
\end{equation}
That is, the truncation by the finite observation time $T$ is not an exponential cut-off but a linear decay with $T-t$.

We test our theory with simulations of a CTRW process. The waiting time PDF was obtained from simulated CTRWs with $\alpha=0.4$ and the observation time $T=60$. 
In Fig.~\ref{fig3}(b), we infer the value of $\alpha$ from the observed data using our theoretical expression Eq.~(\ref{eq_timewindow2}) and using an exponentially truncated power-law 
\begin{equation}
\psi_{\rm ETPL}(t) = \frac{t_r^\alpha}{\Gamma\left(-\alpha,\frac{t_{\rm min}+1}{t_r}\right)-\Gamma\left(-\alpha,\frac{T+1}{t_r}\right)} \frac{{\rm e}^{-(t+1)/t_r}}{(t+1)^{1+\alpha}}.
\label{ETPL}
\end{equation}
Here, $\psi_{\rm ETPL}(t)$ is normalized for $t\in [t_{\rm min};T]$. While apparently both fits explain well the simulation data, the extracted value of $\alpha$ is very different. We confirm that  
our theory Eq.~(\ref{eq_timewindow2}) correctly recovers the original exponent. However, the empirical approach with the exponentially truncated power-law produces an unrealistic estimation for $\alpha$, which was $\alpha=-0.13$ and a cutoff time $t_r=15$.

\subsection{The combined effects of the resolution and time window}\label{secIIc}
Based on our theoretical studies in Sec.~\ref{secIIa} and \ref{secIIb}, here we seek to find the expression of the apparent PDF $\psi_{{\rm obs},S}(t)$ limited by the resolution and time window simultaneously. Combining the two main results Eq.~(\ref{eq_reso2}) and Eq.~(\ref{eq_timewindow}), we find that $\psi_{{\rm obs},S}(t)$ satisfies the following formal expression 
\begin{equation}
\label{eq_combination}
\psi_{{\rm obs},S}(t) = \frac{\psi_{{\rm reso},S}(t)}{\cal N}\int_0^{T-t}d\tau \;P(\tau) \int_0^{T-\tau} du \; \psi_{{\rm reso},S}(u)
\end{equation}
where $\psi_{{\rm reso},S}(t)$ is the inverse Laplace transform of the function given in Eq.~(\ref{eq_reso2}). Unfortunately, it is almost infeasible to obtain the analytic expression of $\psi_{{\rm reso},S}(t)$ for a power-law PDF. Therefore, we have to proceed with an analytical approximation of $\psi_{{\rm reso},S}(t)$ to evaluate Eq.~\eqref{eq_combination} or conduct numerical evaluations of Eq.~\eqref{eq_combination} to get $\psi_{{\rm obs},S}(t)$. 

\begin{figure}
\includegraphics[width=0.9\linewidth]{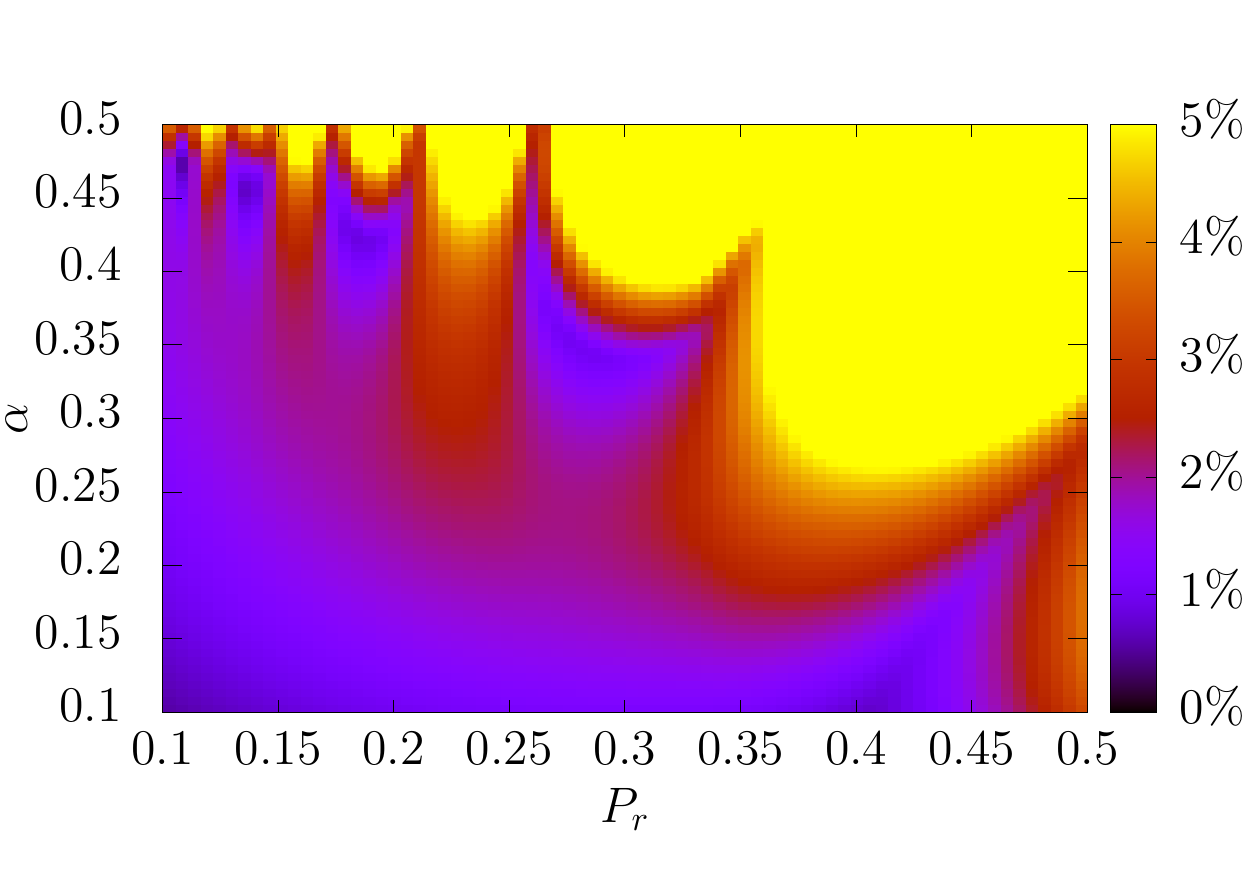}
\caption{Maximum of the relative error of the approximation Eq.~(\ref{eq_reso_approx}) with respect to the numerical inverse Laplace transform of Eq.~(\ref{eq_reso2}) for $t\in [1,100].$}
\label{fig4}
\end{figure}

Let us first proceed in the former method. For the two extreme limits ($t\to0$ and $t\to\infty$), we can separately find the analytic expression of $\psi_{{\rm reso},S}(t)$ for a power-law PDF \eqref{eq:power-law} with $\tau_0=1$. By empirically matching the two limiting results, we end up with the following approximation of  $\psi_{{\rm reso},S}(t)$:
\begin{eqnarray}
\label{eq_reso_approx}
&\overline{\psi}_{{\rm reso},S}(t)\simeq\frac{\psi_S(t)}{1-P_r} +\frac{\alpha(1-P_r)-\frac{\psi_S(0)}{(1-P_r)}}{(t+1)^{2+\alpha P_r}}\\ 
&\nonumber +\sum_{n=2}^{[1/\alpha]}\frac{\alpha^n\Gamma(-\alpha)^n P_r^{n-1}}{(1-P_r)^n\Gamma(-n\alpha)}\left(\frac{1}{(t+1)^{1+n\alpha}} - \frac{1}{(t+1)^{2+\alpha P_r}}\right)\\
&\nonumber +\frac{2P_r\alpha\Gamma(-\alpha)}{(1-P_r)^2(1-\alpha)\Gamma(-\alpha-1)}\left(\frac{1}{(t+1)^{2+\alpha}}-\frac{1}{(t+1)^{2+\alpha P_r}}\right).
\end{eqnarray}

Considering that the approximation is valid if the relative error $\frac{|\psi_{{\rm reso},S}(t)-\overline{\psi}_{{\rm reso},S}(t))|}{\psi_{{\rm reso},S}(t)}<5\%$, we estimate that the domain of validity is such that $P_r+\alpha \simeq 0.65$. In Fig.~\ref{fig4}, we plotted the maximum of the relative error of the approximation Eq.~(\ref{eq_reso_approx}) in the domain $t\in[1,100]$ with respect to the numerical inverse Laplace transform of Eq.~(\ref{eq_reso2}). The heatmap is plotted such that the color displayed is yellow when the relative error is $>5\%$. Therefore, all the remaining area corresponds to the case that the approximation is considered valid.
\begin{figure}
\includegraphics[width=0.49\linewidth]{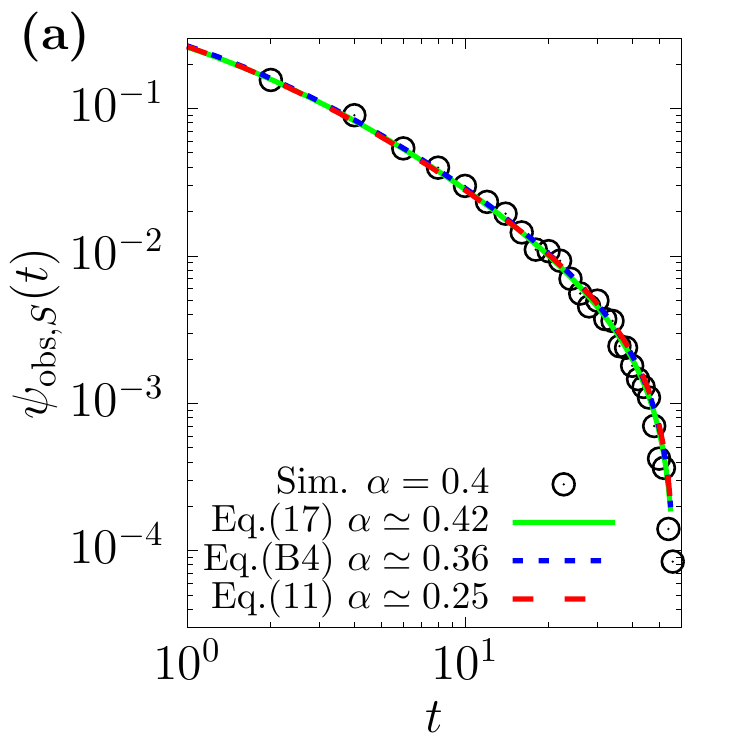}
\includegraphics[width=0.49\linewidth]{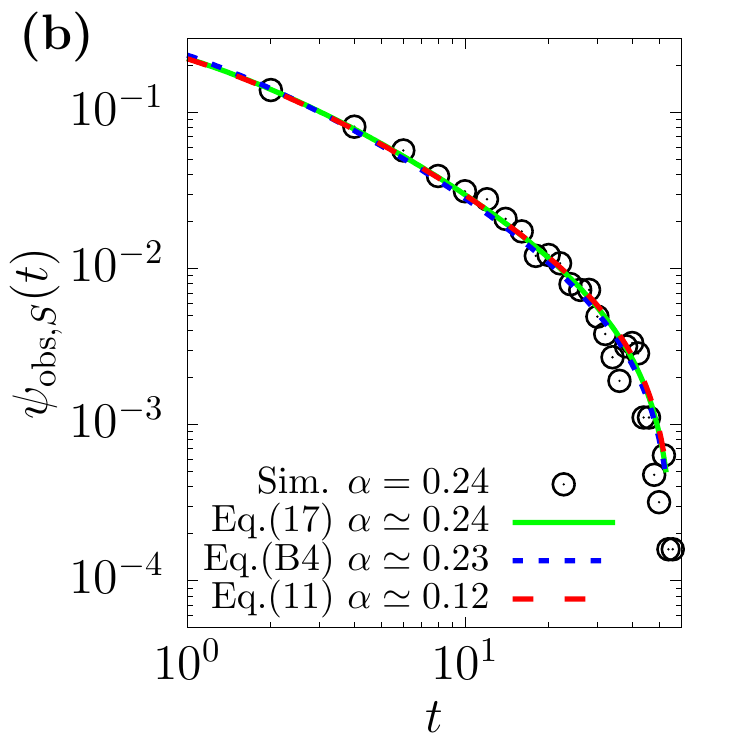}
\caption{The observed waiting time PDF of rest events in the model of a L\'evy walk with rests [Fig.~\ref{fig1}(a)] under the limitation of the resolution and time window. The L\'evy walk process is generated with an exponentially truncated power-law PDF of run times $\psi_\mathrm{run}(t)\propto\frac{{\rm e}^{-t/t_r}}{(t+1)^{1+\eta}}$ and a power-law PDF of rest times $\psi_\mathrm{rest}(t)$ given by Eq.~\eqref{eq:power-law}. See the schematic of this process in Fig.~\ref{fig1}(a). The simulation parameters are:
(a) $\eta = 0.1$, $t_r=20$, $\alpha = 0.4$, and $r\Delta t=1$. (b) $\eta = 0.4$, $t_r=20$, $\alpha = 0.24$, and $r\Delta t=0.6$.
In both panels, the symbols represents the simulations results. The solid line shows the fit using our exact theory Eq.~(\ref{eq_combination2}) while the dotted line is the fit using the approximated PDF, $\overline{\psi}_{{\rm obs},S}(t)$, from Eq.~\eqref{eq_appendix_approx}. The dashed line represents the fit using $\psi_{{\rm time},S}(t)$ [Eq.~(\ref{eq_timewindow2})] that incorporates the effect of time window in the absence of the resolution limitation.  }
\label{fig5}
\end{figure}
We note that the approximation Eq.~(\ref{eq_reso_approx}) consists of a sum of power-law terms; therefore replacing each power-law by their corresponding expression Eq.~(\ref{eq_timewindow2}), we obtain the final approximated expression for the combined effects $\psi_{\rm obs,app}$ given in the Appendix Eq.~(\ref{eq_appendix_approx}). 

Alternatively, we can obtain $\psi_{{\rm obs},S}(t)$ by numerically evaluating the formal expression  Eq.~(\ref{eq_combination}). For this, we rewrite the integral expression in Eq.~(\ref{eq_combination}) as
\begin{eqnarray}
\label{eq_combination2}
\psi_{{\rm obs,}S}(t) &=& \frac{{\cal L}^{-1}[\hat{\psi}_{{\rm reso},S}](t)}{\cal N}\\ \nonumber
& & \times \left({\cal L}^{-1}\left[\frac{\hat{\psi}_{{\rm reso},S}}{s^2}\right](T)-{\cal L}^{-1}\left[\frac{\hat{\psi}_{{\rm reso},S}}{s^2}\right](t) \right).
\end{eqnarray}
We then use the Gaver-Stehfest algorithm \cite{Steh1970} to numerically evaluate the inverse Laplace transform in the above expression. 

We test our theory with an example of L\'evy walk with rests schematically explained in Fig.~\ref{fig1}(a), which will be our dynamic model in the next section for the application of the developed theories to an experimental system. The simulated process and its event time PDFs can be understood as those of a CTRW process with the $P_r$ defined in Eq.~\eqref{eq:P_rCTRW}. In Fig.~\ref{fig5}, we simulate a L\'evy walk with rests where the run (ballistic phase) event is generated with random sojourn times governed by a truncated power-law and the rest event with random waiting times governed by a power-law PDF. For further information, see the Appendix A for the simulation detail. Here, we extract the (power-law) rest time PDFs from the simulated trajectories in the presence of the resolution and time window limits.   

Shown in Fig.~\ref{fig5} are the observed rest time PDFs of the L\'evy walk process for two distinct parameter cases. To infer the power-law exponent $\alpha$ we fit the simulation data with the following three theoretical expressions ($S=\textrm{"rest"}$): (i) the expected PDF $\psi_{{\rm obs,}S}(t)$ based on the exact theory Eq.~(\ref{eq_combination2}), (ii) the approximated PDF $\overline{\psi}_{{\rm obs},S}(t)$ via Eq.~(\ref{eq_appendix_approx}), and (iii) $\psi_{{\rm time},S}(t)$ based on  Eq.~(\ref{eq_timewindow2}), which is the theoretical PDF that only takes into account the effect of time window. The results show that although the data seems to be fitted well with the three expressions the fitting values greatly differ. It is demonstrated that the exact expression (i) estimates the underlying exponent successfully for the two data set [(a) \& (b)] even though the observation time window is not sufficiently long. Using the approximated PDF $\overline{\psi}_{{\rm obs},S}(t)$, we can infer $\alpha$ in good agreement with the value in the simulation when the resolution is small (Fig.~(\ref{fig5}b)). However, the estimation becomes bad as the resolution is larger (Fig.~(\ref{fig5}a)). Finally, we confirm that without incorporating the resolution effect the PDF $\psi_{{\rm time},S}(t)$ fails the correct estimation of the power-law exponent.

\section{An Experimental application: the transport of mRNP particles in neuronal cells}\label{secIII}
\begin{figure}
\begin{center}
\includegraphics[width=0.48\linewidth]{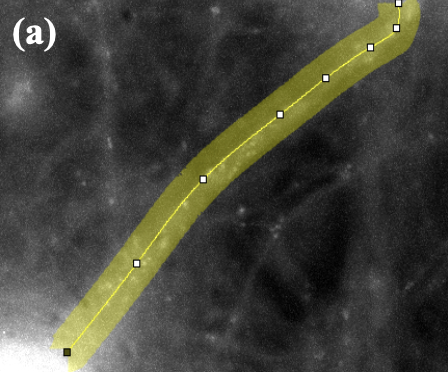}
\includegraphics[width=0.48\linewidth]{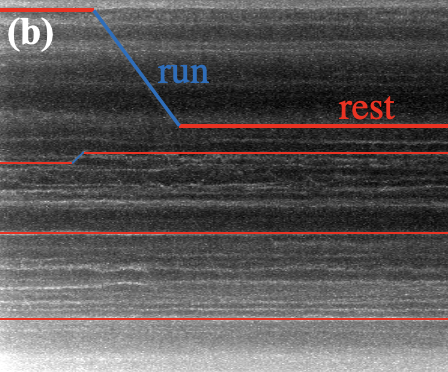}
\par\vspace{0.3cm}
\includegraphics[width=0.90\linewidth]{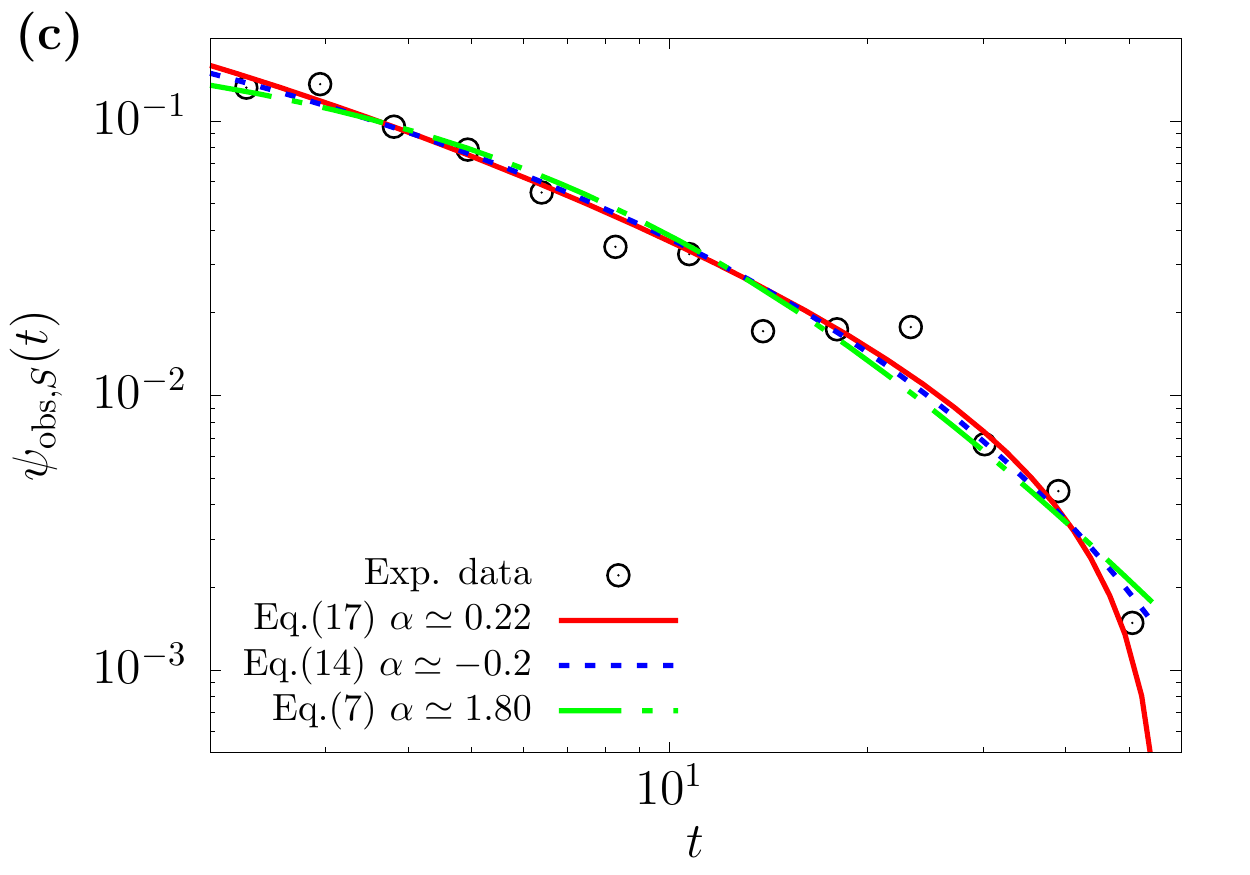}
\end{center}
\caption{ (a) Live-cell image of a hippocampal neuron showing fluorescently labeled $\beta$-actin mRNPs. Images were taken every time interval $\Delta t=200~ {\rm ms}$. (b) A typical kymograph for an ensemble of mRNP particles spotted along an observed dendrite as in (a). The single mRNP motion is composed of run and rest phases, which are marked in blue and red colors, respectively. (c) Distribution of the duration of rest events $\psi_{\rm rest}(t)$. Symbols: Experimental PDF. Solid line: the best fit using our theory Eq.~(\ref{eq_combination2}) with $\alpha\approx0.22$. 
Dashed line: the best fit using an ETPL Eq.~(\ref{ETPL}) with $\alpha\approx0.2$ and $t_b\approx23.1$~s. The resolution has been estimated to be such that $P_r\simeq 0.45$. Dashed-dotted line: the best fit using a simple power-law Eq.~(\ref{eq:power-law}) with $\alpha\approx1.80$ and $\tau_0\approx12.27$. The power-law function was normalized to unity on the range of the plot.}
\label{fig6}
\end{figure}

As an experimental application of our theory, we determined the statistics of the dynamics of $\beta$-actin mRNP complexes transported along the dendrites of neurons by motor proteins [Fig.~\ref{fig6}(a)]. Previously, we have shown that the motion of $\beta$-actin mRNP complexes consists of an alternation of rests and runs. See the kymograph of mRNP particles in Fig.~\ref{fig6}(b).
We performed experiments following a similar protocol used in \cite {Song2018} except for the use of bicuculline for neuronal stimulation. Briefly, we cultured hippocampal neurons from the MCPxMBS mice \cite {park2014}, in which every single endogenous $\beta$-actin mRNAs are labeled with multiple green fluorescent proteins. At 14--16 days \emph{in vitro}, we stimulated the neurons by treating them with 50 $\mu$M bicuculline for 20 min. At 40--60 min after the onset of the stimulation, we imaged the movement of individual $\beta$-actin mRNP particles in proximal dendrites (0--50 $\mu$m from the cell body) at 200 ms interval over a one-minute time course. Being fluorescently labeled, the individual mRNPs appear as bright spots in Fig.~\ref{fig6}(a). Kymographs of the time-lapse images were generated, from which the position of fluorescently labeled mRNPs were detected and registered, see Fig.~\ref{fig6}(b). 

As observed in the kymograph, the stochastic diffusion dynamics of single mRNP particles are described by the L\'evy walk with rests introduced in Fig.~\ref{fig1}(c). The motion of mRNPs is an alternating dynamics of the run and rest phases. Here, the run is a ballistic movement with random sojourn times while the rest is the stop state with waiting times governed by a PDF distinct from the run's.       
After the identification of the trajectories $\{X_i, i = 0\ldots N\}$ from the kymographs, we proceeded to the determination of their states. For this, we calculated the  velocity $\frac{X_{i+1}-X_i}{\Delta t}$ where $\Delta t=0.2~{\rm s}$ was the time interval of imaging. After averaging and filtering this velocity profile, we determined the dynamic state of the mRNP particle by applying the following criterion: if the velocity was $<v_{\rm threshold} = 0.3 \;{\rm \mu m/s}$ then the particle was considered in a rest state, if not it was considered in a run. 

We now focus on the PDF $\psi_{\rm rest}$ of the waiting time of the rest event. For the run event, it was shown that the sojourn time PDF was not of a power-law~\cite{Song2018}, which was explained well by an exponentially truncated power-law $\psi_\textrm{run}(t)\sim \frac{{\rm e}^{-t/t_r}}{(t+1)^{1+\eta}}$ with $\eta=0.52$ and $t_r = 12.5$~s. Using this information, we calculated the probability $P_r$ that a run is too short to be detected. Estimating that the resolution is about $n_{\rm min}=5$ data points or about $1~{\rm s}$, it gives $P_r = \int_0^{n_{\rm min}\Delta t}\psi_\textrm{run}(t)\;\textrm{d} t=1-\frac{\Gamma(-\eta,(n_{\rm min}\Delta t+1)/t_r)}{\Gamma(-\eta,1/t_r)}\simeq 0.45$.

In Fig.~\ref{fig6} we plot the rest time PDF from our experiment. We infer the power-law exponent $\alpha$ of the PDF from the data with $\psi_{{\rm obs,}S}(t)$ that takes into account the combined effects of the resolution and time window [Eq.~\eqref{eq_combination2}]. For comparison, we also measure the best fit values of $\alpha$ with power-law PDFs, Eq.~\eqref{ETPL} (exponentially truncated power-law) and
Eq.~\eqref{eq:power-law} (power-law).
We find that the fit with the single power-law results in $\alpha\approx1.8$, while the fit with the exponentially truncated power-law (ETPL) gives $\alpha\approx-0.2$ and the truncation characteristic time $t_b = 23.1$. Although the experimental data are explained well, at least visually, by both fit curves, the two reference methods give inconsistent values for $\alpha$. Moreover, neither of them turns out to be physically correct in  that the $\beta$-actin mRNPs motion follows an aged L\'evy walk from our previous study~\cite{Song2018}. 
Namely, the L\'evy walk with rest events governed by the above ETPL or by a power-law PDF with $\alpha(\approx1.8)$ of $>1$ cannot age due to the finite first moment of the rest times. Therefore, the fitted exponents $\alpha$ using the two reference power-law PDFs have to be rejected for dynamical reasons. 
Contrary to these empirical approaches, we obtain $\alpha \approx0.22$ from our theory based on $\psi_{{\rm obs,}S}(t)$. The estimated value not only explains well the data over the entire time window (red curve in Fig.~\ref{fig6}) but also falls within $0<\alpha<1$ to be compatible with the reported aging transport dynamics of the $\beta$-actin mRNP particle. We note that    $\alpha \approx 0.22$ is smaller than the value $\alpha\approx 0.32$ obtained in the case of non-stimulated neurons experiment \cite{Song2018}. 

To sum up, in this work, we have developed a general framework to properly infer the distribution of event times when the data is subjected to experimental constraints: time-spatial resolution and time window. We have derived the observed distribution for both constraints taken separately Eqs.~\eqref{eq_reso2} \& \eqref{eq_timewindow2} and combined them to obtain the main result of the paper Eq.~(\ref{eq_combination2}). We have applied this theory to a practical example and showed how to use our expressions to obtain the correct exponent of a power-law PDF. It is noteworthy to mention that while we have presented the experimental constraints affecting temporal observables, e.g. the distribution of the duration of the events, it is immediately applicable for spatial variables. Also, even though we have focused on power-law distributions throughout this paper, the formal expressions Eqs.~\eqref{eq_reso2}, \eqref{eq_timewindow}, and \eqref{eq_combination2} are applicable for any distributions.\\

\begin{acknowledgments}
This work was supported by the National Research Foundation (NRF) of Korea, No.~2020R1I1A1A0107179) (X.D.), No.~2020R1A2C2007285 (H.Y.P.), and No.~2020R1A2C4002490 (J.-H.J.).
\end{acknowledgments}

\appendix
\section{Simulation Details}
In Fig.~\ref{fig5}, we have simulated the process of a L\'evy walk with rests such that the rest time PDF is $\psi_{\rm rest}(t) = \frac{\alpha}{(t+1)^{1+\alpha}}$ and the run time PDF is $\psi_{\rm run}(t) \sim \frac{{\rm e}^{-t/t_r}}{(t+1)^{1+\eta}}$. It consists of an alternation of run and rest whose respective random durations are drawn from the corresponding PDFs. At time $t=0$, all trajectories start with a run. The runs are set to be ballistic motions at a constant velocity $v=1$. The random times governed by the two PDFs are generated using the inverse transform sampling method \cite{Devr86}. We simulated $N=10^5$ trajectories consisting of points spaced by $\Delta t=0.2$ within the observation time window $[\tau_a, ~\tau_a+T]$ where $\tau_a=100$ and $T=60$. 

Once a trajectory is simulated, it is recorded with a given resolution parameter $r$ in the following way. Let us assume that a L\'evy walk with rests is simulated with a variation of event times as schematically illustrated in Fig.~\ref{fig_S1}(a). For every event in $\mathcal{S}$ and $\mathcal{S'}$ states, if $t_{\rm event}>r\Delta t$, this state is recognized to occur and the event is recorded with a duration of $t_{\rm event}$. However, if $t_{\rm event}<r\Delta t$ (e.g., in $\mathcal{S}$ state), this event fails to be detected. Then, the event (in $\mathcal{S}$ state) is recorded as a part of the previous event (in $\mathcal{S}'$ state) and, accordingly, the latter state has an increased duration time by $t_{\rm event}$. We repeat the same protocol for the next events, obtaining the recorded trajectory as in Fig.~\ref{fig_S1}(b). 

\setcounter{figure}{0}
\renewcommand{\figurename}{FIG.}
\renewcommand{\thefigure}{S\arabic{figure}}
\begin{figure}
\includegraphics[width=0.95\linewidth]{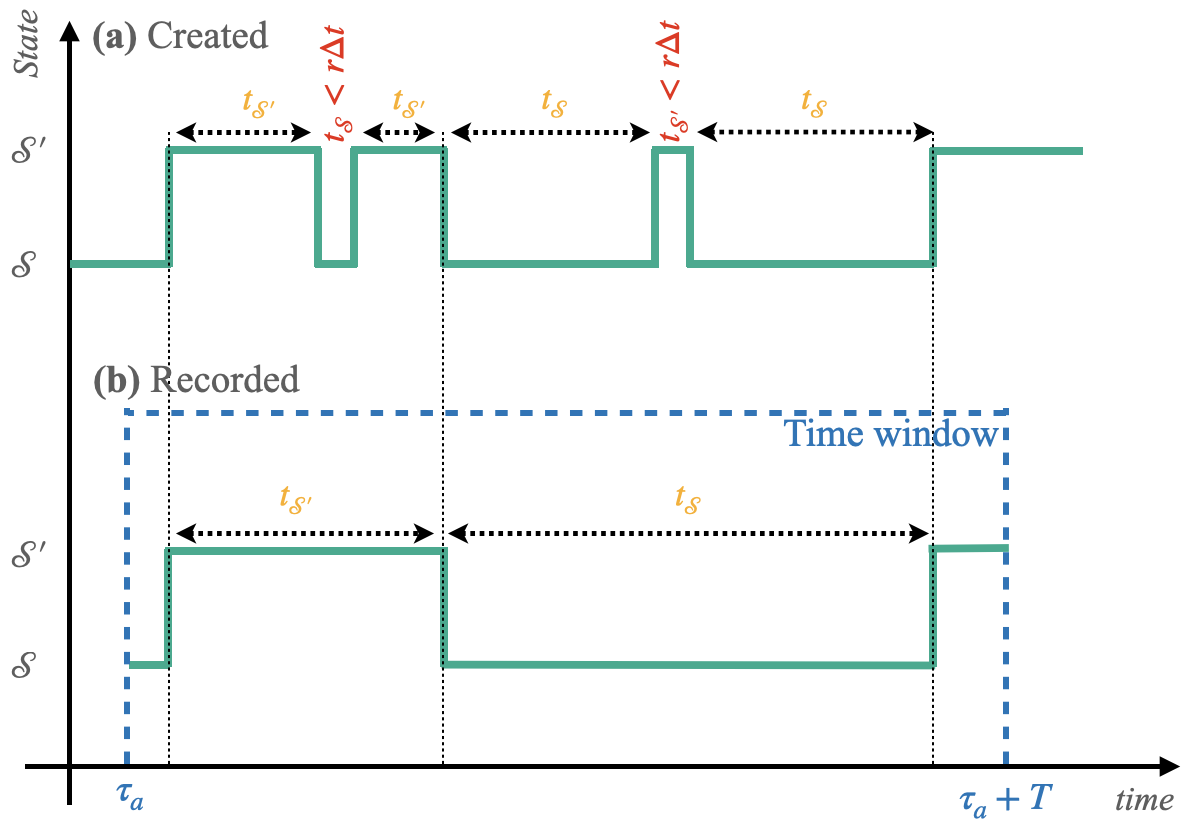}
\caption{Example of sampled trajectories. During the sampling, if a duration $t_{S,S'}$ drawn from the corresponding distributions is smaller than $r\Delta t$, the change of event will not appear in the recorded trajectory. a) the trajectory created before taking care of time window or resolution. b) the observed or the recorded trajectory.}
\label{fig_S1}
\end{figure}


\section{Derivation of Eq.~\eqref{eq_reso_approx}}
The expression of Eq.~\eqref{eq_reso_approx} is obtained phenomenologically by matching the initial behavior and the asymptotic behavior of  Eq.~(\ref{eq_reso2}).
The series expansion of Eq.~(\ref{eq_reso2}) gives the large-time behavior
\begin{eqnarray}
    \overline{\psi}_{\rm asympt}(t)&\simeq \frac{1}{1-P_r}\left(\psi_S(t) + \sum_{n=2}^{[1/\alpha]}\frac{\alpha^n\Gamma(-\alpha)^n P_r^{n-1}}{(1-P_r)^n\Gamma(-n\alpha)}\frac{1}{(t+1)^{1+n\alpha}} \right. \nonumber\\ 
& \left.+\frac{2P_r\alpha\Gamma(-\alpha)}{(1-P_r)^2(1-\alpha)\Gamma(-\alpha-1)}\frac{1}{(t+1)^{2+\alpha}}\right).
\end{eqnarray}
The asymptotic limit of Eq.~(\ref{eq_reso2}) gives the $t\rightarrow 0$ limit 
\begin{equation}
    \overline{\psi}_{\rm init} (t) = \alpha(1-P_r). 
\end{equation}
We construct the final expression by requiring that it satisfies both limits and by imposing that it only contains power-law terms
\begin{equation}
    \overline{\psi}_{{\rm reso},S}(t)=\overline{\psi}_{\rm asympt}(t) -\frac{\overline{\psi}_{\rm asympt}(0)}{(t+1)^{2+\alpha P_r}} +\frac{\overline{\psi}_{\rm init}(t)}{(t+1)^{2+\alpha P_r}},
\end{equation}
which gives Eq.~(\ref{eq_reso_approx}). 

To obtain the expression for the combined effects $\overline{\psi}_{{\rm obs},S}(t)$, we replace every simple power-law that appears in Eq.~(\ref{eq_reso2}) by the expression for the time window effect  Eq.~(\ref{eq_timewindow})
\begin{eqnarray}
\label{eq_appendix_approx}
&\overline{\psi}_{{\rm obs},S}(t)\simeq\frac{\psi_S(t)}{1-P_r}\left[T-t+\frac{(T+1)^{1-\alpha}-(t+1)^{1-\alpha}}{\alpha-1} \right] \nonumber \\
&+\frac{\alpha(1-P_r)-\frac{\psi_S(0)}{(1-P_r)}}{(t+1)^{2+\alpha P_r}}\left[T-t+\frac{(T+1)^{-\alpha P_r}-(t+1)^{-\alpha P_r}}{\alpha P_r} \right]\\ 
&\nonumber +\sum_{n=2}^{[1/\alpha]}\frac{\alpha^n\Gamma(-\alpha)^n P_r^{n-1}}{(1-P_r)^n\Gamma(-n\alpha)}\frac{1}{(t+1)^{1+n\alpha}}\left[T-t+\frac{(T+1)^{1-n\alpha}-(t+1)^{1-n\alpha}}{n\alpha-1} \right]\\
&\nonumber -\sum_{n=2}^{[1/\alpha]}\frac{\alpha^n\Gamma(-\alpha)^n P_r^{n-1}}{(1-P_r)^n\Gamma(-n\alpha)}\frac{1}{(t+1)^{2+\alpha P_r}}\left[T-t+\frac{(T+1)^{-\alpha P_r}-(t+1)^{-\alpha P_r}}{\alpha P_r} \right]\\
&\nonumber +\frac{2P_r\alpha\Gamma(-\alpha)}{(1-P_r)^2(1-\alpha)\Gamma(-\alpha-1)}\frac{1}{(t+1)^{2+\alpha}}\left[T-t+\frac{(T+1)^{-\alpha}-(t+1)^{-\alpha}}{\alpha} \right]\\
&\nonumber -\frac{2P_r\alpha\Gamma(-\alpha)}{(1-P_r)^2(1-\alpha)\Gamma(-\alpha-1)}\frac{1}{(t+1)^{2+\alpha P_r}}\left[T-t+\frac{(T+1)^{-\alpha P_r}-(t+1)^{-\alpha P_r}}{\alpha P_r} \right].
\end{eqnarray}

\bibliography{powerlaw}
\bibliographystyle{apsrev4-1}

\end{document}